\documentclass[12pt]{futurenetworkandmobilesummit10}

\usepackage{fancyhdr}
\usepackage{lastpage}
\usepackage{graphicx}
\usepackage{floatflt}
\newcommand{\comments}[1]{}

\fancypagestyle{plain}{
  \setlength{\topmargin}{-7mm}
    \setlength{\headheight}{77pt}

    \fancyhead[c]{
        \fontsize{9pt}{10pt}
        \it
        \begin{flushleft}
        \hspace*{25mm} Future Network and MobileSummit 2010 Conference Proceedings \\
        \hspace*{25mm} Paul Cunningham and Miriam Cunningham (Eds) \\
        \hspace*{25mm} IIMC International Information Management Corporation, 2010 \\
        \hspace*{25mm} ISBN: 978-1-905824-16-8 \\
        \end{flushleft}
    }
}

\def\ind{{\mathchoice {\rm 1\mskip-4mu l} {\rm 1\mskip-4mu l}
{\rm 1\mskip-4.5mu l} {\rm 1\mskip-5mu l}}}

\usepackage{amsmath,amssymb,amscd,amsthm}
\newtheorem{lemma}{Lemma}
\newtheorem{theorem}{Theorem}
\newtheorem{definition}[theorem]{Definition}

\newcommand{\ul}{\underline}

\newcommand{\T}{\mathcal{T}}

\newcommand{\mc}{\mathcal}

\newcommand{\ds}{\displaystyle}

\title{Coverage games in small cells networks}
\author{M. Le Treust, H. Tembine, S. Lasaulce, and M. Debbah
}

\begin{document}

\maketitle

\begin{abstract}
This paper considers the problem of cooperative power control in distributed small cell wireless networks. We introduce a novel framework, based on repeated games, which models the interactions of the different transmit base stations in the downlink. By exploiting the specific structure of the  game, we show that we can improve the system performance by selecting the Pareto optimal solution as well as reduce the price of stability.
\end{abstract}


\section{Introduction}

In this paper we consider the problem of cooperative power control
in distributed wireless networks. In such networks, transmitters can
decide their control policy freely. Because of multi-user
interference the decision of the different transmitters are
inter-dependent, which makes game theory a natural paradigm to study
the problem of distributed power control. As the power control game
take several times (e.g., because a transmitter sends several
packets), repeated games are considered. In repeated games, selfish
transmitters can have interest to cooperate. This framework is very
relevant for the downlink of small cells networks
where small base stations have not much information to implement
cooperative power control policies. It is realistic to assume that
small base stations are connected but only by low-capacity links.
Therefore, repeated games where only low signalling between the
players are a
suited framework to address the problem under investigation.

Repeated games with complete information (every player knows the set
of players, all the action spaces, all the payoff functions) are
known to have several outcomes. A well known result in that field is
the so-called {\it Folk Theorem} \cite{Aum81,Sorin92} which
characterizes the set of equilibrium payoffs if the players have
perfect monitoring (every player is able to
observe at each stage the actions chosen by all the other players), every
feasible and individually rational payoff (at least the minmax point) can be
 obtained by an equilibrium strategy of the repeated game.
Whereas the knowledge of {\it perfect monitoring} can be acquired in
certain scenarios where appropriate estimation and sensing
mechanisms are implemented, our goal here is to show that some of
these information assumptions can be relaxed by exploiting the
specific structure of the networking game. This says in particular
that one specific operating point, a specific Pareto optimal
solution or a {\it global optimum} can be approximated by  repeated
equilibrium play under suitable assumptions. As a consequence, many
results on inefficiency of equilibria in static games can be
examined in repeated game setting and the performance can be
improved. To this end, we consider a proximity graph of monitoring which allow us
to relax the full observation assumptions  on the other players.



Our contribution is to provide explicit conditions over the coverage game such as implement an optimal action plan for the long-run game inspired from the work of Renault Tomala (1998 \cite{RenaultTomala98}) : any deviation of a player is followed by an identification procedure which will isolate the eventual deviator and an appropriate punishment plan.
One of the main consequence of this result is the possibility to select a specific operating point, namely a Pareto optimal solution (such as global optimum, Bargaining solution) which can improve the system performance, and hence reducing the {\it price of stability} (PoS) (the gap between the best equilibrium payoff and the optimal social welfare).
Note that Nash equilibria, Wardrop equilibria, Stackelberg solutions can be suboptimal and inefficient in generic static games. Under proximity graph of strategic observation, the repeated game approach can lead to the global optimum if the initial plan is this operating point (see Theorem \ref{TheoremOptStrat}). In contrast to most of learning algorithms that try to reach ``equilibrium'' of the static game, here we examine equilibria of the long-run game which can be global optimum for the static game. In particular the price of stability  is one.

The rest of the paper is structured as follows. In Section~ \ref{secnmodel}, we present the network model. In Section~\ref{oneshot}, we analyze the one-shot game between small base stations (SBSs). In Section~\ref{secrepeated}, we study the repeated coverage game.

\section{Network model} \label{secnmodel}

We consider $S$ small base stations (SBS) and a large number of
mobile stations. The SBS index is denoted by $i \in \{1,...,S\}$.
They SBS are assumed to exploit the same frequency band. Around each
SBS, mobile stations are assumed to be distributed geographically in
a plane according to a density $\lambda_i(x,y)$ where
$(x,y)\in\mathbb{R}^2$ correspond to the  coordinates of a mobile
station located at the position $(x,y).$ The density $\lambda_i(x,y)=0$ outside a range of a certain radius $R_i$ from SBS $i.$  We use the polar coordinate representation: the radius from the origin is $r=(x^2+y^{2})^{\frac{1}{2}},\ $ the angle $\theta\in [0,2\pi[$ is determined by $\cos \theta= \frac{x}{r},\ \sin \theta= \frac{y}{r}$ for $r\neq 0,$. The relevant channel model to describe this
situation is the interference channel. 

In the
scenario under investigation, the interference channel comprises $S$
transmitters and a large number of receivers. Let $l_1,\ldots,l_S$ be the location of the SBSs,  $l_i=(l_{i1},l_{i2},l_{i3})\in \mathbb{R}^3$ with $l_{i3}\neq 0.$ We assume that  the $l_i\neq l_j,\ \forall\ i\neq j.$ We denote by $d:\ \mathbb{R}^3 \times \mathbb{R}^3\longrightarrow \mathbb{R}_{+}$ the  Euclidean distance in $\mathbb{R}^3.$

$$
d((x,y,z),(x',y',z')):=\left((x-x')^2+(y-y')^2+(z-z')^2 \right)^{\frac{1}{2}}.
$$

Let  $d_i(x,y):=d(l_i,(x,y,0))$ be the Euclidian distance from SBS $i$ to a MS located at the position $(x,y).$
$$
d_i(x,y):=d(l_i,(x,y,0))=\left((x-l_{i1})^2+(y-l_{i3})^2+l_{i3}^2 \right)^{\frac{1}{2}}\geq |l_{i3}|>0.
$$
 The transmit power of
SBS $i$ is denoted by $p_i$. If an MS is located at $(x,y)$ the
downlink SINR associated with SBS $i$ is given by
\begin{equation}
\mathrm{SINR}_i(x,y) = \frac{g_{ii}(x,y) p_i  }{\sigma^2 +
\ds{\sum_{j\neq i} g_{ji}(x,y) p_j}}
\end{equation}
where $g_{ii}(x,y)=a_{ii}(x,y) d_i(x,y)^{-\gamma_i}$ represents the channel gain (path loss) of the link
between SBS $i$ and an MS located in the point $(x,y)$,
$g_{ji}(x,y)=a_{ji}(x,y)d_j(x,y)^{-\gamma_j}$ represents the channel gain (path loss) of the link between SBS $j$ and an MS located in the point $(x,y)$,\ $\gamma_j$ is a real number $\gamma_j>2.$

\subsection*{Remark} The mapping $(x,y)\longmapsto \mathrm{SINR}_{i}(x,y)$ has a finite limit at $(0,0).$
  We can choose the function $a_{ii}(x,y)$ into the form $b_{ii} d_i(x,y)^{-\gamma_i} \ind_{\{ d_i(x,y)\leq R_i \}}$ (No interference from $i$ outside a range of a maximum radius $R_i$)
Similarly,
$$a_{ji}(x,y)=b_{ji} d_j(x,y)^{-\gamma_j} \ind_{\{ d_j(x,y)\leq R_j \}},$$
$b_{ii}$ and $b_{ji}$ are
    positive constants.


\section{One-shot coverage game definition} \label{oneshot}

The one-shot coverage game is defined by the triplet
\begin{equation}
\mc{G} = \left(\mc{S}, \{\mc{P}_i\}_{i\in\mc{S}}, \{u_i\}_{i\in\mc{S}}
\right).
\end{equation}
The set $\mc{S}=\{1,...,S\}$ is the set of players who are the SBS,
the set $\mc{P}_i =[0, p_i^{\max}]$ is the action space of SBS $i$,
and the utility function of SBS $i$ is defined as follows:
\begin{equation}\label{Definition:utilityfunction}
u_i(p_1,p_2,...,p_S) = \int \int_{\mathbb{R}^2}
\lambda_i(x,y) \log_2 \left(1 + \frac{g_{ii}(x,y) p_i }{\sigma^2
+ \ds{\sum_{j\neq i} g_{ji}(x,y) p_j}} \right)  \mathrm{d}x\ \mathrm{d}y\end{equation}
Assume that $g_{ii}(x,y)$ has the above form. Then, $u_i(p_1,p_2,...,p_S)$ can be rewritten as
\begin{equation}
\int_{0}^{R_i} \int_{0}^{2\pi}
\lambda_i(r\cos\theta,r\sin\theta) \log_2 \left(1 + \frac{g_{ii}(r\cos \theta,r \sin \theta) p_i }{\sigma^2
+ \ds{\sum_{j\neq i} g_{ji}(r\cos \theta,r \sin \theta) p_j}} \right)  r\mathrm{d}r\ \mathrm{d}\theta
\end{equation}

\subsection{Existence and uniqueness of the one-shot Nash equilibrium}
The players are rational and the above description is common knowledge (every player is rational know $\mc{G}$ and every player know that every player is rational and know $\mc{G}$ and so on). An important game solution
concept is the Nash equilibrium, i.e., a point from which no player has interest in
unilaterally deviating.
An action profile $P$ is a pure Nash equilibrium of $\mc{G}$  if
\begin{equation}\forall i \in \mc{S},\ \forall\ p_i'\in \mc{P}_i,\ u_i(p_1,\ldots,p_{i-1},p_{i},p_{i+1},\ldots,p_S)\geq u_i(p_1,\ldots,p_{i-1},p'_{i},p_{i+1},\ldots,p_S)
\end{equation}

We say that a  strategy for a player is dominated if there exists another strategy for her which is better for her no matter what choices the opponents make.
Since such strategies are suboptimal, the player would eliminate dominated strategies. This dominance analysis reduces the set of possible outcome of the game. In some games, a dominance analysis leads to a unique prediction of the outcome when players are rational. We say that these games are said {\it dominance solvable}.

The next lemma shows that our one-shot coverage game is  a dominance solvable game.
\begin{lemma}
The one-shot game is a dominance solvable game.
\end{lemma}

 As a corollary, the game $\mc{G}$ has a unique Nash equilibrium given by $P^{NE}=(p_1^{\max},\ldots,p_S^{\max}\label{power:NE}).$ The minmax level is determined by $P^{NE}$. Denote $u_i^{NE}=u_i(P^{NE})$ the utility of SBS $i$ at the Nash equilibrium.
\subsection{Inefficiency of NE}
 If there is no interfering ranges between the  SBSs then NE is a global optimum (maximum sum utility).
For interfering ranges of base stations, the Nash equilibrium is clearly inefficient compared to the global optimum of of the long-run interaction. For example, for the long-term interaction, a time-sharing solution would be better than the NE.
Next, we define a specific Pareto optimal solution, namely a bargaining solution (Nash bargaining, Rubinstein bargaining, Kalai-Smorodinsky bargaining etc). The Kalai-Smorodinsky (denoted KS) bargaining \cite{ks1975,ks2} is one the bargaining model which has a very nice geometric interpretation of the problem of ``splitting a cake''. The  KS bargaining is a fair  solution in the sense that the minimum utility is maximized. Moreover, in the long-run game, the KS bargaining solution achieves the convex hull of the the feasible utilities and strictly dominate the one-shot Nash equilibrium.

\begin{figure}[htb]
\centering
  \includegraphics[width=12cm,height=15cm]{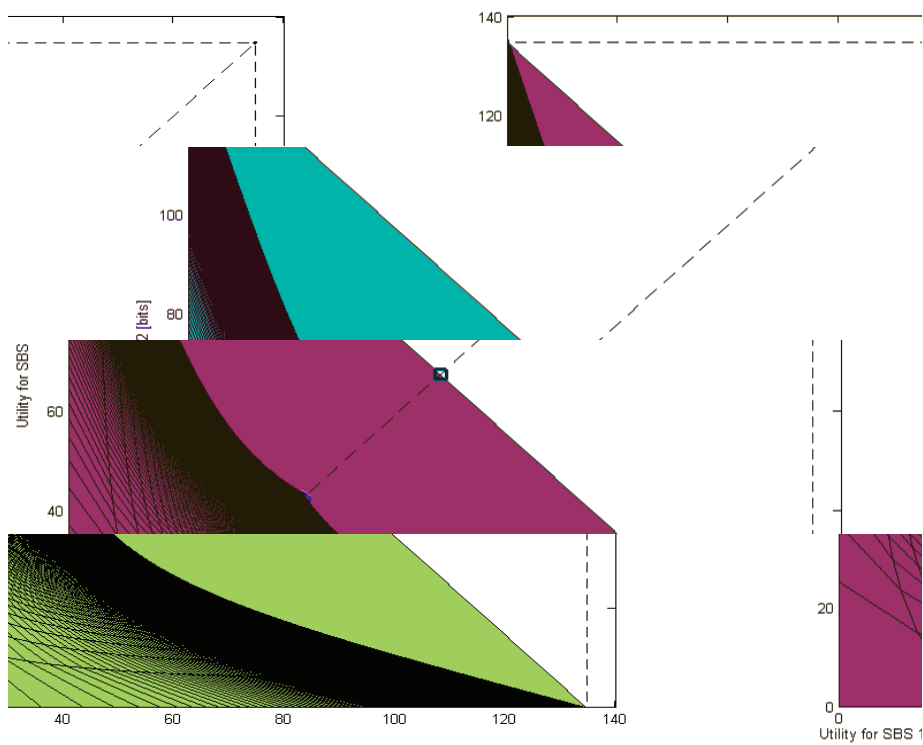}\\
  \caption{Utility region of coverage repeated game. }\label{figks}
\end{figure}

By definition, the KS bargaining solution corresponds to the maximum point of the region of convex hull of feasible utilities that is located on the line joining a disagreement point (for example the NE $u^{NE}=(u_i^{NE})_{i \in \mc{S}}$)\label{utility:NE} and the maximum utility profile denoted $\bar{u}=(\bar{u}_i)_{i \in \mc{S}}$ where $\bar{u}_i=\max_{\underline{p}}u_i(\underline{p}$\label{utility:max}). Denote $p^{KS}=(p_i^{KS})_{i \in \mc{S}}$\label{power:KS} the power vector that achieves the KS bargaining solution and $u^{KS}=(u_i^{KS})_{i \in \mc{S}}$\label{utility:KS} the corresponding utility profile. In Figure \ref{figks}, we consider a simple example for which the KS bargaining solution consist in an fair time sharing. In next section, we develop a repeated approach leading to the KS bargaining solution.

%

\section{Repeated coverage game } \label{secrepeated}

Assume the SBS are synchronized (say at the frame level) and update
their transmit power every frame. We consider a discounted repeated
game played over a large number of frames.
 Taking into account the past behavior of the player allow to construct Pareto-optimal equilibrium strategies. The assumption that the players observe the actions of all the others players, at the end of each stages seems unrealistic in our network game.
 We suppose here the players are able to sense their environments to detect the power transmission level of their neighbor. We model this situation using a graph of strategic observation, saying that a SBS $i$ observes the action power of each SBS neighbor $j\in G(i)$ in the graph $G$. Strategic signaling is an essential assumption to guarantee a robust equilibrium condition on the proposed strategic action plan.\\

The repeated game is denoted $\Gamma=(\mc{S},(\T_i)_{i \in \mc{S}},(\vartheta_i)_{i \in \mc{S}},(s_i)_{i \in \mc{S}})$, where $\mc{S}$ is the set of players, $(\T_i)_{i \in \mc{S}}$ is the set of strategies, $(\vartheta_i)_{i \in \mc{S}}$ is the vector of the long-term payoffs and $s_i$ is the strategic observation function of player $i$ : $s_i : \mc{P} \rightarrow S_i$ defined by the observation graph $G$. If action $p\in \mc{P}$ was played in the last stage, player $i$ will received a strategic signal $s_i(p)$ which disclose the information about his neighbor's actions. From now on, we assume that, for a given transmission block or game stage $t \geq
1$, each device $i$ knows and take into account the actions that
have been played by his neighbors $G(i)$ in the past, before choosing his stage action $p_i(t)$. We denote by $p^i(t)$ the sequence of actions
played before time $t$ and observed by player $i$: $
p^i(t) = (p_i(t), (p_j(t-1))_{j \in G(i)}).$
The vector $\ul{h}^i_t = (p^i(1),...,p^i(t-1))$
is called the private history of player $i$ at
time $t$ and lies in the set $
\mc{H}^i_t  = \left(\bigotimes_{j \in G(i)} \left[0, p_j^{\mathrm{max}}
\right] \right)^{t-1}.$

\begin{definition}[Players' strategies in the RG] \emph{A pure strategy for
player $i \in \mc{S}$ is a sequence of functions $\left(\tau_{i,t}
\right)_{t \geq 1}$ with}
\begin{equation}
\label{eq:strategy-rg} \tau_{i,t}: \left|
\begin{array}{ccc}
\mc{H}^i_t & \rightarrow & \mc{P}_i\\
 \ul{h}^i_t & \mapsto & p_i(t).
\end{array}
\right.
\end{equation}
\end{definition}
The strategy of player $i$ will therefore be denoted by $\tau_i$
while the vector of strategies $\ul{\tau} = (\tau_1, ..., \tau_{\mc{S}})$
will be referred to a joint strategy. A joint strategy
$\ul{\tau}$ induce in a natural way a unique plan of action  $(p(t))_{t\geq
1}$. To each profile of powers $p(t)$ corresponds a certain
instantaneous payoff $u_i(p(t)) $ for player $i$. In our
setup, each player does not care about what he gets on a given block
but what he gets over the whole duration of the game. This is why we
consider a payoff function resulting from averaging over the
instantaneous payoff. In order to quantify the fact that the
transmitters can value short-term and long-term gains differently we
use the model of infinitely repeated games with discounting
\cite{Aum81}. The averaged payoff for player $i$ can
then be defined as follows.

\begin{definition}[Players' payoffs in the RG] \emph{Let $\ul{\tau} = (\tau_1, ..., \tau_{\mc{S}})$
be a joint strategy. The payoff for player $i \in \mc{S}$ is
defined by:}
\begin{equation}
\label{eq:utility-rg} v_i(\ul{\tau}) = \sum_{t \in \mathbb{N}^*}
\lambda (1 - \lambda)^{t-1} u_i(\ul{p}(t))
\end{equation}
\emph{where $\ul{p}(t)$ is the power profile of the action plan
induced by the joint strategy $\ul{\tau}$.}
\end{definition}
The parameter $ 0< \lambda <1$ is the discount factor which is seen as the game stopping probability \cite{Sorin92}: the probability that the
game stops at stage $t$ is thus $\lambda (1-\lambda )^{ t-1 } $.
This shows that the discount factor is also useful to study wireless
games where a player enters/leaves the game.

At this point, equilibrium strategies in the repeated game can be
defined.
\begin{definition}[Equilibrium strategies in the RG] \emph{A joint strategy $\ul{\tau}$ supports an equilibrium of
the repeated game defined by $\Gamma=(\mc{S},(\T_i)_{i \in \mc{S}},(\vartheta_i)_{i \in \mc{S}},(s_i)_{i \in \mc{S}})$  if}
\begin{equation}
\forall i \in \mc{S}, \forall \tau_i', \ v_i(\ul{\tau}) \geq
v_i(\tau_1, ..., \tau_{i-1}, \tau_i',\tau_{i+1},...,\tau_{\mc{S}})
\end{equation}
\end{definition}

\subsection{Strategic signal on a Proximity Graph of Base Station} \label{secmain}

The strategic observation structure is a fundamental in order to define an optimal action plan and guarantee a robust equilibrium condition at each time of the game duration. In order to guarantee the  cooperation on an optimal operating point (as the KS bargaining solution), the players SBS should be able to detect any deviating behavior, identify the SBS which deviate and start a punishment mechanism. Inspired from \cite{RenaultTomala98}, we first propose a repeated game strategy leading to the KS bargaining solution. Second we show that this strategy also satisfy a robust equilibrium condition on for each time of the game duration.
\begin{definition}\label{opt-strat}
We propose a strategy that lead to a Pareto-optimal utility vector defined by the KS bargaining, such that non deviation from this cooperative plan will be profitable.
\begin{itemize}
\item The cooperative plan take place from the first stage of the game and consist in playing the KS bargaining power profile $p^{KS}  \ref{power:KS}$ as long as no deviation is detected.
\item If a player deviate, then begin the procedure of the sets of suspects.
\item After the end of the procedure, start an adequate punishment plan targeted at the deviator.
\end{itemize}
\end{definition}

\textbf{Procedure of the set of suspects}

Using the results of Renault Tomala (1998 \cite{RenaultTomala98}), we define  the procedure of the set of suspects which spread the identity of the deviator among all the players in a finite number of stages. To implement this strategy, the graph of strategic signalling should satisfy the following 2-connectivity property.
\begin{definition}
A graph $G$ is 2-connected if $G$ is connected and for each $i\in \mc{S}$, the graph $G^i$ (when player $i$ has been removed) is connected.
\end{definition}
We give an explicit description of the procedure of the set of suspect. We refer to \cite{RenaultTomala98} for detailed proofs and intuitive examples.
First divide the stages into block of length $l=\log(2^{\mc{S}})$. Each subset $N\subset \mc{S}$ of player is mapped into a sequence of actions $p(1),...,p(l)$ which is publicly known. The sequences of power level are used to encode every subset of players in order to communicate between the players. If during some block $m$, the neighbor $j$ of player $i$ do not follow the main plan. Then, player $i$ elaborate a set of suspect including player $j$ and other possible deviating neighbors.  At the beginning of block $m+1$, he play the sequence of moves corresponding to his set of suspect.
Now, from block to block, the players confront their set of suspect with their neighbors, adding the new suspects and excluding innocents players. The 2-connectivity property is essential to innocent all the players except the deviator. The date at which a player $k$ enter in the set of suspect of player $i$ and of player $j$ is predictable, knowing the structure of the graph. If player $k$ did not deviate, then another player, say $i$, will remark an incoherence between the date of the set of suspect including $k$ and the length of the shortest paths from $k$ to $i$. Then player $i$ exonerate player $k$ so as his neighbor in the next block. The identity of the deviator become common knowledge after $ \bar{n}= l\cdot \max(1,2l-5)$ stages. \\

\textbf{Punishment Plan}
Once the procedure of the set of the suspects finish, every SBS know the identity of the deviating SBS $i\in \mc{S}$.
The punishment plan of the strategy consist for the neighbor $G(i)$ of player $i$ in playing a one-shot Nash equilibrium $p^{NE}$ until the end of the game whereas the other players continue to play the optimal KS bargaining $p^{KS}$ action power (see Fig. 1).\\

Such a strategy is proved to lead to an optimal operating point such that no deviation from the above algorithm could be profitable.
\begin{theorem}\label{TheoremOptStrat} If $G$ is 2-connected, the discount factor satisfy the following condition
\begin{eqnarray}
\forall i\in \mc{S},\quad \lambda < 1 - \sqrt[\bar{n}]{\frac{\bar{u_i}-  u_i^{KS} }{\bar{u_i} - u_i^{NE}}}
\end{eqnarray}
Then the above strategy (\ref{opt-strat}) is an equilibrium strategy.\\
\end{theorem}
Note that any other feasible utility vector characterized by the Folk theorem could be considered here.

\begin{proof}\label{proof-theorem}
Suppose a player $i$ deviates, thus the procedure of the suspect spread his identity among the players and the punishment phase push down his payoff under $u_i^{NE}$ until the end of the game. Compare the total deviation payoff :
\begin{eqnarray*}
  \sum_{s=1}^{\bar{n}} \lambda(1-\lambda)^{s-1} \bar{u_i} + \sum_{s\geq \bar{n}+1} \lambda(1-\lambda)^{s-1} u_i^{NE} &<& \sum_{s=1}^{\bar{n}} \lambda(1-\lambda)^{s-1} u_i^{KS} + \sum_{s\geq \bar{n}+1} \lambda(1-\lambda)^{s-1}  u_i^{KS}\\
 &\Longleftrightarrow & (1-(1-\lambda)^{\bar{n}})(\bar{u_i} -\tilde{u_i}) <  (1-\lambda)^{\bar{n}} ( u_i^{KS} -  u_i^{NE})\\
&\Longleftrightarrow& \lambda < 1 - \sqrt[\bar{n}]{\frac{\bar{u_i}-  u_i^{KS} }{\bar{u_i} - u_i^{NE}}}
\end{eqnarray*}
The recursive structure of the discounted utility function implies that this inequality does not depends on a particular stage.
This condition, over the stop probability $\lambda$ insures that the equilibrium condition is valid for every player in every stage of the game, thus the proposed  strategy is an equilibrium strategy of the repeated game.\\
\end{proof}

\section{Numerical Results} \label{secnumericalresult}
Our theoretical result are illustrated with a valid simple model of two SBS. In figure \ref{figks} we derive the achievable utility region for our two base stations considering the utilities defined by equations \ref{Definition:utilityfunction} where the densities $\lambda_i(x,y)$ are supposed to be uniform and the channel gains $g_{ii}(x,y)$ are constant over the range of SBS $i$. This figure depicts the utility region and it's convex hull with the Nash equilibrium utility and the KS bargaining utility. The repeated game approach improved the utility of both SBS.

\section{Concluding remark} \label{secconclusion}
We have studied a proximity-based network coverage games and have shown that the KS bargaining utility can be achieved for our model of discounted repeated game. This result is general and could be applied in a larger class of communication game over a fixed network. It would be nice to investigate an application of the suspect's procedure to classical web-protocols. From the game theoretical point of view, it would be interesting to extend this result to sub-game perfect equilibrium or to relax the information hypothesis or to introduce a stochastic graph.

\bibliographystyle{ieeetr}
\bibliography{BiblioMael}

\begin{thebibliography}{1}

\bibitem{Aum81}
R.~Aumann, ``Survey of repeated games,'' essays in game theory and mathematical
  economics in honor of oskar morgenstern, edited by V. Bohm, Bibliographisches
  Institut, Mannheim, 1981.

\bibitem{Sorin92}
S.~Sorin, {\em Repeated Games with Complete Information, in Hanbook of Game
  Theory with Economic Applications}, vol.~1.
\newblock Elsevier Science Publishers, 1992.

\bibitem{RenaultTomala98}
J.~Renault and T.~Tomala, ``Repeated proximity games,'' {\em International
  Journal of Games Theory}, vol.~27, pp.~539--559, 1998.

\bibitem{ks1975}
E.~Kalai and M.~Smorodinsky, ``Other solutions to nash's bargaining problem,''
  {\em Econometrica}, vol.~43, pp.~513--518, 1975.

\bibitem{ks2}
E.~Kalai, ``Solutions to the bargaining problem,'' {\em in Hurwicz L., D.
  Schmeidler and H. Sonnenschein (eds.), Social Goals and Organization,
  Cambridge University Press, Cambridge.}, 1985.

\end{thebibliography}

\end{document}